\documentclass[11pt]{article}
\usepackage{fullpage,doublespace}
\usepackage[margin=1in]{geometry}
\usepackage{amssymb, amsthm, amsmath}
\usepackage{graphicx,stackengine}
\usepackage[authoryear]{natbib}
\usepackage{bm}
\usepackage{lineno}
\usepackage{times}
\usepackage{placeins}
\usepackage{float}
\usepackage[ruled,vlined]{algorithm2e}
\usepackage{comment}
\usepackage[lofdepth,lotdepth]{subfig}
\usepackage[export]{adjustbox}

\newcommand{\btheta}{ \mbox{\boldmath $\theta$}}

\newcommand{\bbeta}{ \mbox{\boldmath $\beta$}}

\newcommand{\be}{ \mbox{\bf e}}

\newcommand{\bx}{ \mbox{\bf x}}
\newcommand{\bX}{ \mbox{\bf X}}
\newcommand{\bB}{ \mbox{\bf B}}

\newcommand{\by}{ \mbox{\bf y}}
\newcommand{\bY}{ \mbox{\bf Y}}

\newcommand{\bV}{ \mbox{\bf V}}

\newcommand{\iid}{\stackrel{iid}{\sim}}

\newcommand{\beq}{ \begin{equation}}
\newcommand{\eeq}{ \end{equation}}
\newcommand{\beqn}{ \begin{eqnarray}}
\newcommand{\eeqn}{ \end{eqnarray}}

\begin{document} 

\begin{center}
{\Large Monitoring Deforestation Using Multivariate Bayesian Online Changepoint Detection with Outliers}\\\vspace{6pt}
{\large Laura Wendelberger\footnote[1]{Department of Statistics, North Carolina State University}, Josh Gray\footnote[2]{Center for Geospatial Analytics, North Carolina State University}, Brian J Reich$^1$, Alyson Wilson$^1$}\\
\today
\end{center}

\begin{abstract}
\noindent Near real time change detection is important for a variety of Earth monitoring applications and remains a high priority for remote sensing science. Data sparsity, subtle changes, seasonal trends, and the presence of outliers make detecting actual landscape changes challenging. \cite{Adams2007} introduced Bayesian Online Changepoint Detection (BOCPD), a computationally efficient, exact Bayesian method for change detection. Incorporation of prior information allows for relaxed dependence on dense data and an extensive stable period, making this method applicable to relatively short time series and multiple changepoint detection. In this paper we conduct BOCPD with a multivariate linear regression framework that supports seasonal trends. We introduce a mechanism to make BOCPD robust against occasional outliers without compromising the computational efficiency of an exact posterior change distribution nor the detection latency. We show via simulations that the method effectively detects change in the presence of outliers. The method is then applied to monitor deforestation in Myanmar where we show superior performance compared to current online changepoint detection methods. 
\vspace{12pt} 
\end{abstract}

\section{Introduction}\label{s:intro}
Online change detection is increasingly necessary for monitoring remote sensing data as it becomes available~\citep{Woodcock2020}. In global monitoring applications, remote sensing data containing multiple spectral bands are collected over time for large geographical areas. Timely detection and characterization of deforestation, heavy construction, and other land use changes are necessary in order to respond to environmental or security threats. We aim to build a multivariate screening algorithm to flag changed areas of interest for further analysis and inspection.

Modern change detection methods use massive amounts of individual remotely sensed images to construct time series covering large spatial scales. Thus, automated methods of identifying unusable observations, primarily clouds and their shadows, are necessary. However, failures in cloud masking algorithms do occur \citep{Zhu2012a} and lead to the presence of outliers in the data stream.

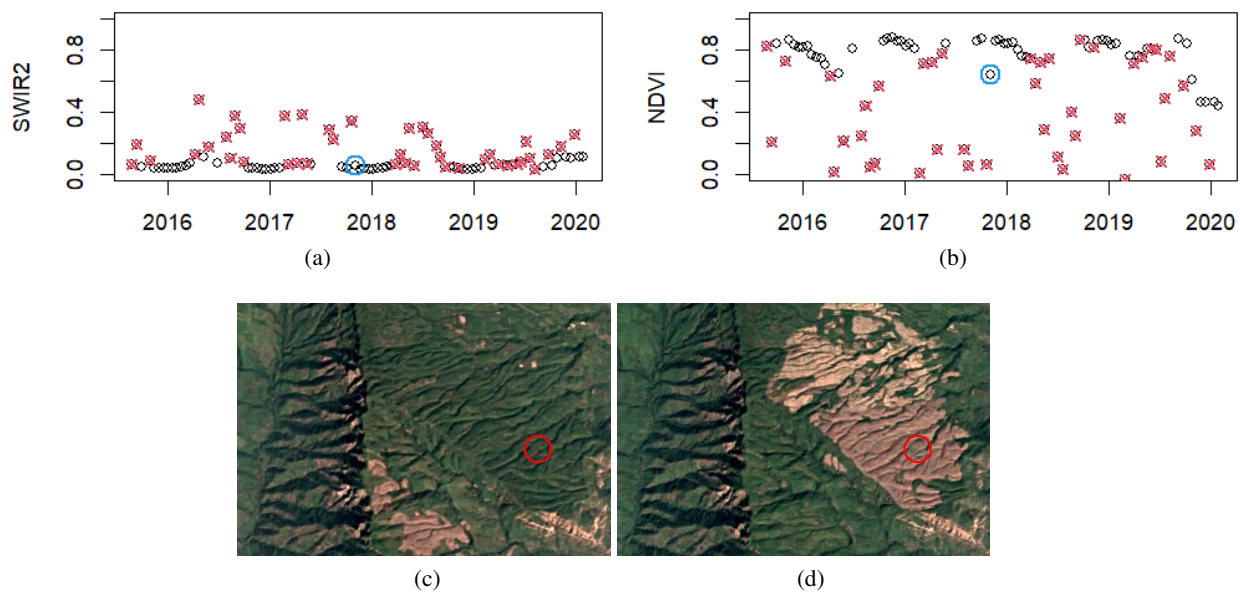
\begin{figure}[h]
\centering
 \subfloat[][]{
   \includegraphics[width=0.5\textwidth]{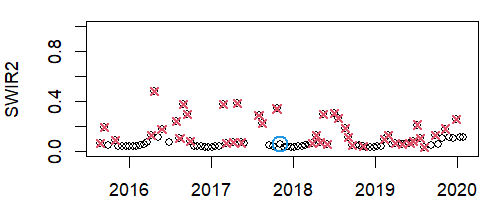}
 }
 \subfloat[][]{
   \includegraphics[width=0.5\textwidth]{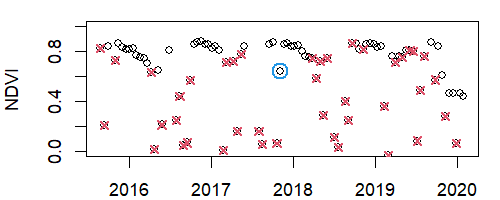}
   }\\
  \subfloat[][]{\includegraphics[width=0.3\textwidth,trim={2cm 4cm 20cm 5cm}, clip]{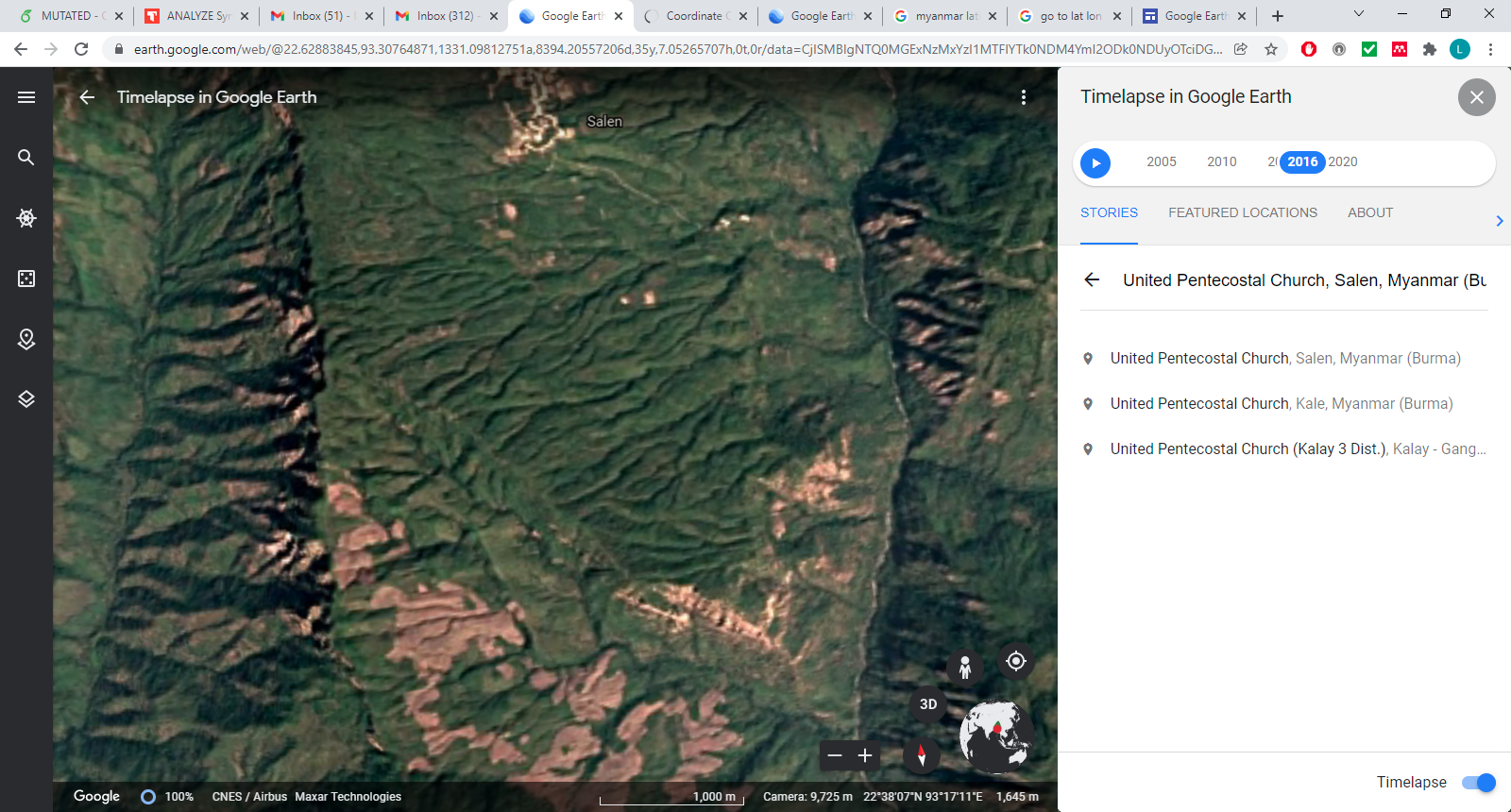}
    \begin{picture}(0,0)
  \thicklines
  \pdfliteral{1 0 0 rg}
  \put(-30,40){\circle{10}}
  \pdfliteral{0 0 0 rg}
  \end{picture}}
  \subfloat[][]{\includegraphics[width=0.3\textwidth,trim={2cm 4cm 20cm 5cm}, clip]{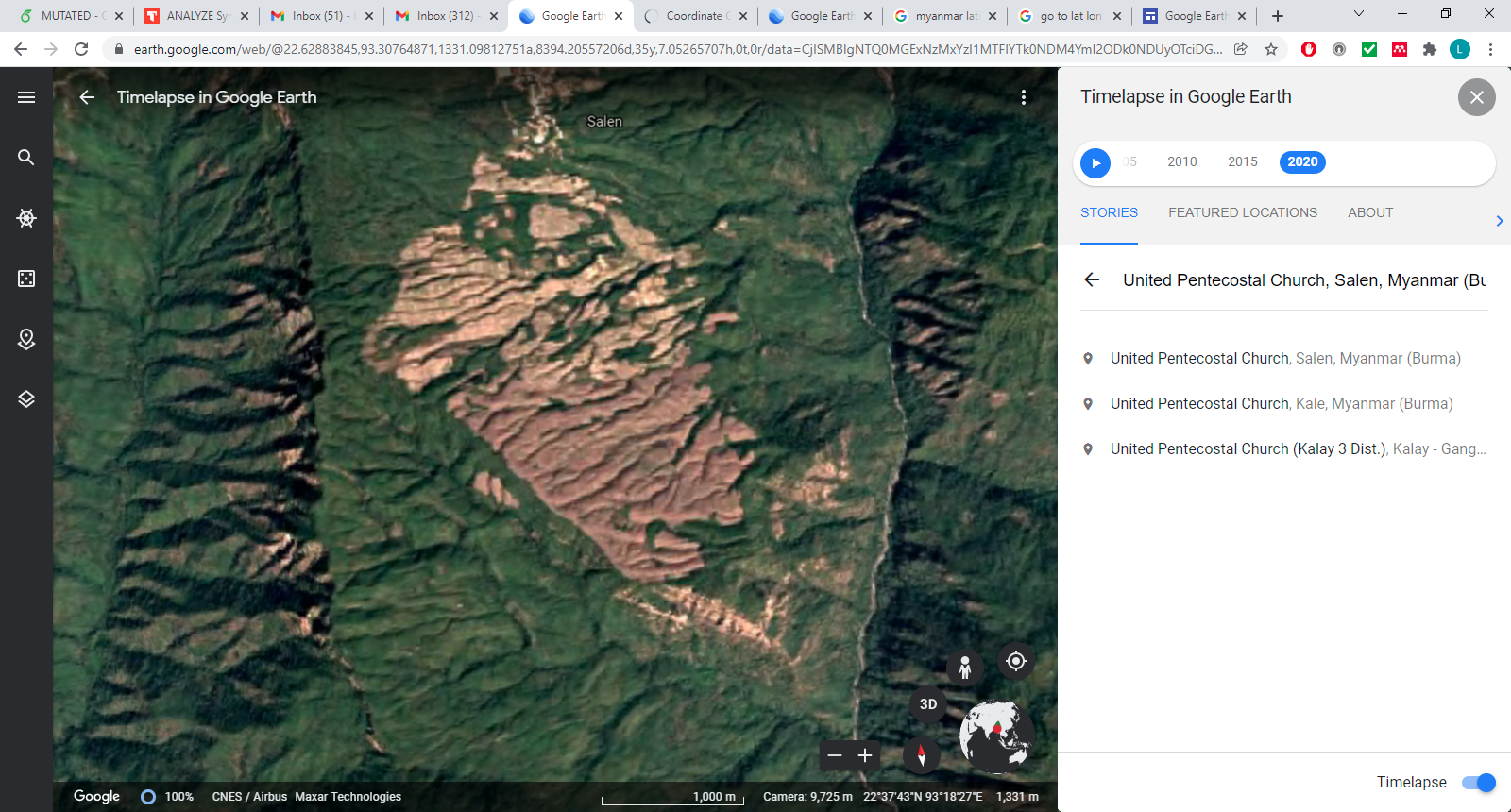}
  \begin{picture}(0,0)
  \thicklines
  \pdfliteral{1 0 0 rg}
  \put(-30,40){\circle{10}}
  \pdfliteral{0 0 0 rg}
  \end{picture}
  }
 \caption{Time series of the unitless indices SWIR2 (a) and NDVI (b) with filtered points (red X's) and an unfiltered outlier (blue circle) show long term change beginning Oct 24, 2019. Google Earth imagery from Myanmar in 2016 (c) and 2020 (d) visualizes the land cover change for the location of interest (circled in red)}
  \label{fig:myanmar_data_site}
\end{figure}

As an example, consider a forest disturbance in Myanmar shown in Figure \ref{fig:myanmar_data_site}. For a single pixel in the region, the associated data streams have much of the data eliminated based on cloud and aerosol filtering, resulting in a relatively sparse signal to monitor. Seasonal patterns explain much of the signal variation within each year. There is a real event on October 24th, 2019, but there is an undetected anomaly on November 3, 2017. A good monitoring algorithm must be robust to transient changes like seasonal variation and outliers. 

A multitude of retrospective as well as online detection algorithms have been developed over the last four to five decades. ``Offline" methods look at an entire time series retrospectively, using all available data to identify changepoints throughout the series. Hypothesis-test-based methods compare model parameters under a single model versus one with a break~\citep{Chow1960,Hinkley1970,Bai1998,Bai2003}. A natural extension to multiple break detection balances a cost function for the models with the number of breaks, as in binary segmentation~\citep{Scott1974} and its variation wild binary segmentation~\citep{Fryzlewicz2014}. The Bayesian time series forecasting method Prophet~\citep{Taylor2018} supports common time series attributes like seasonality and delivers changepoints as a byproduct of its forecasts. Using an offline method repeatedly for online data monitoring can be computationally intensive and hypothesis-test based methods in particular require consideration of multiple-testing. Further, many of these offline methods are not designed for monitoring multidimensional data.

On the other hand, online methods analyze data as it is observed and must update their analyses as each new piece of data comes in. Considerations for both storage and computational time are more prominent for online methods than offline ones. In the field of remote sensing, Continuous Change Detection and Classification (CCDC; \citeauthor{Zhu2014}, \citeyear{Zhu2014}) is a common method used to flag change based on repeated large deviations from the expected output based on a stable period and then classify the type of land cover change. Since this paper is focused only on detection of the existence of change, we examine only the Continuous Change Detection (CCD) component of the algorithm. Modifications to CCD incorporate cumulative sums (CUSUM; \citeauthor{Bullock2020}, \citeyear{Bullock2020}), similar to Breaks for Additive Season and Trend (BFAST; \citeauthor{Verbesselt2010a}, \citeyear{Verbesselt2010a,Verbesselt2010,Verbesselt2012}). A limitation of these methods is that both CCD, its variants, and BFAST are dependent on a long stable period to learn model parameters before applying those models during a monitoring period. Verifying the stability of the training period presents its own challenges and, even when the training period is stable, the large amounts of data associated with long training periods may be difficult to store and access at scale. Spatial considerations based on signal within a moving window can help filter out false change~\citep{Lin2020}, but can be heuristic and may struggle where a site does not change all at once.  Another limitation is that in a multivariate setting, CCD assumes independent signals.

Bayesian Online Changepoint Detection (BOCPD) was simultaneously introduced in \cite{Adams2007} and \cite{Fearnhead2007}. Unlike previous Bayesian methods which are more focused on retrospective segmentation, BOCPD uses a message-passing algorithm to calculate a posterior distribution for the most recent changepoint. Online hyperparameter optimization \citep{Turner2009,Wilson2010,Caron2011} improves BOCPD performance. In a similar approach, \cite{Reiche2015} proposed a Bayesian land cover change classification method. This classification view is limited by dividing possible land cover into discrete categories and requires training data to characterize class-specific distributions. \cite{pmlr-v80-knoblauch18a} extended BOCPD to incorporate model uncertainty regarding the form of the model into BOCPD with Model Selection (BOCPDMS). They also set up a spatiotemporal model framework, adding dependence on a neighborhood into the models for each location. This method flags change of an entire image, but is not set up to identify nor localize partial changes within an image, particularly when those changes are a small proportion of the overall scene, and offers no mechanism for partially missing data neither spatially nor temporally. BOCPD is susceptible to high false detection rates in the presence of outliers. \cite{NEURIPS2018_a3f390d8} incorporates General Bayesian Inference (GBI) with $\beta$-divergences to reduce the influence of outliers on the analysis, consequently reducing the mislabeling of outliers as changepoints. The robustness to outliers and consideration of model uncertainty come at the cost of increased computation time. \cite{Fearnhead2018} use efficient dynamic programming algorithms~\citep{Maidstone2016} to minimize a bounded loss function, which are less sensitive to outliers compared to squared-error loss. BOCPD with hyperparameter optimization was the best performing multivariate changepoint detection method in a comprehensive review of changepoint algorithms using numerous benchmark datasets~\citep{Burg2020}.

We propose a multivariate Bayesian changepoint detection methods that is robust to outliers. Our approach is based on the assumption that outliers arise from a completely different generating function than the data does. Drawing on concepts from BOCPDMS~\citep{pmlr-v80-knoblauch18a}, consider that there is some uncertainty over whether data comes from the process of interest or from another source. Given an outlier distribution, it is possible to efficiently calculate the posterior probability that any given point is an outlier and weight its influence accordingly so that they neither cause a false detection nor distort estimated models of the process. Consideration of a such a large number of possible models is unwieldy, so approximations that only trigger outlier detection for recent suspected points are introduced. Likely outliers can be completely removed from the data \textit{in situ}.

The remainder of the paper proceeds as follows. The motivating data are described in Section \ref{s:data}. Section \ref{s:multivariate_changepoint_model} sets up notation for the multivariate changepoint model and reviews BOCPD. Section \ref{s:outlier_detection_and_removal} introduces the outlier detection framework and approximations for implementing it online. Section \ref{s:changepoints} details our rule for classifying changepoints. Computation is discussed in Section \ref{s:comp}.  The method is evaluated using a simulation study in Section \ref{s:sim} and applied to areas undergoing deforestation in Myanmar in Section \ref{s:app}.  Section \ref{s:discussion} concludes.

\section{Deforestation in Myanmar}\label{s:data}
Automated forest monitoring enables quick identification and response to sudden deforestation events. NASA produces and distributes well registered, terrain corrected surface reflectance products from Landsat 8 radiance observations at a nominal resolution of 30 meters \citep{Vermote2016}. Normalized difference vegetation index (NDVI; \citeauthor{Tucker1979}, \citeyear{Tucker1979}) is a unitless red and near-infrared based index commonly used to measure vegetation properties. Landsat 8 also provides unitless observations in a portion of short wave infrared spectrum that are sensitive to water, particularly vegetation moisture (SWIR2). NDVI and SWIR2 are thus complementary signals useful for monitoring changes of interest on Earth, and are used throughout the paper as indicators of latent land cover change. The values in the Quality Assessment (QA) band denote whether a pixel contains cloud, cirrus, cloud-shadow, water or snow. It is usually a poor assumption to have confidence in NDVI and SWIR2 values for pixels at times where they contain these ephemeral components.

The Global Land Analysis \& Discovery (GLAD; \citeauthor{Hansen2016}, \citeyear{Hansen2016}) 2019 Forest Alert Data was used as a starting point for the data assemblers (Ian McGregor and Natalie Chazal [North Carolina State University]) to search the vicinity of suspected disturbance locations using high resolution, commercial PlanetScope \citep{PlanetTeam} imagery. Analysts identified the location and temporal window for forest disturbances. Landsat 8 data associated with the coordinates was recorded and the date of disturbance was annotated. The date of disturbance is the first date at which imagery shows that the land has undergone sustained change, i.e., deforestation has begun. Depending on the temporal density of the PlanetScope images and Landsat 8 data, evidence of the disturbance may be present in Landsat 8 data at an earlier date than it is first identified in imagery. The date of disturbance, taking this uncertainty into account, will be used to evaluate the performance of change monitoring algorithms on real data.

\begin{figure}
    \centering
    \includegraphics[width=\textwidth]{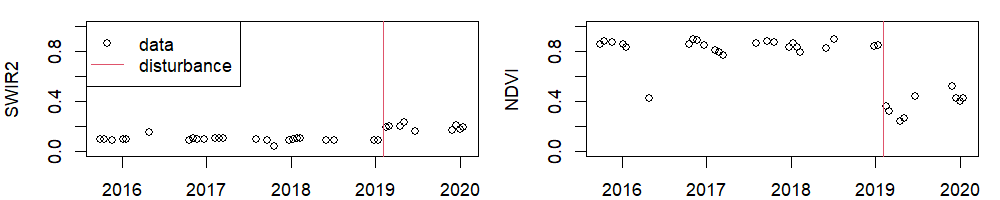}
    \caption{Observed SWIR2 and NDVI measurements of a $30\times 30$m location in Myanmar with a deforestation event identified on Feb 2, 2019.}
    \label{fig:myanmar_data}
\end{figure}

The dataset of interest contains time series, as shown in Figure~\ref{fig:myanmar_data}, for $114$ pixel locations in Myanmar from July 4, 2015 to January 30, 2020. Since clouds, atmospheric conditions, and precipitation are not indicative of sustained land cover change, it is best to exclude these phenomena from the analysis, so any data whose QA value indicates the presence of any of these obstructions is eliminated from consideration and treated as missing. Pixels with radiometric saturation or aerosol values that were not ``low aerosol" were eliminated for similar reasons.

\section{Multivariate changepoint model}\label{s:multivariate_changepoint_model}
In this section we review the Bayesian method for multivariate change-point detection that does not consider outliers. This is extended to handle outliers in Section \ref{s:outlier_detection_and_removal}.
\subsection{Bayesian linear regression model}\label{s:bayesian_linear_regression_model}
Let $\bY_t = [Y_{t1},...,Y_{td}]^T$ be the observation at time step t.  In our analysis, $Y_{tj}$ is the data for index $j$ on day $t$ and $d=2$ indices (NDVI and SWIR2).  We will denote the data from days $s$ to $t$ as $\bY_{s:t} = [\bY_s,\bY_{s+1},...,\bY_t]^T$. Similarly, let $\bX_t = [X_{t1},
\hdots,X_{tk}]^T$ be the covariates at time $t$ and the covariates from days $s$ to $t$ are denoted as $\bX_{s:t}=[\bX_s,\bX_{s+1},\hdots,\bX_t]^T$. At time $t$, there is a partition of the data into states $g_t \in \{1,2,\hdots\}$, as in a Product Partition Model \citep{Barry1993}. The changepoints between states are $\bm{c}=\{c_0,c_1,c_2,\hdots\}$ with $c_0=0$ so that state $h$ occurs between changepoints $c_{h-1}$ and $c_h$. Denote the state at time $t$ as $g_t=h$ so that $t \in [c_{h-1},c_h)$. A multivariate Bayesian linear regression model is chosen to model multi-dimensional data about a linear trend and account for its correlation structure within each state,
\begin{equation}\label{eq:likelihood}
    \bY_t \vert \bbeta_h,\bm{\Sigma}_h,g_t=h \sim \text{Normal}(\bX_t \bbeta_h, \bm{\Sigma}_h),
\end{equation}
where parameters $\bbeta_h$ and $\bm{\Sigma}_h$  are $k\times d$ and $d\times d$, respectively. The unknown state parameters $\btheta_h = [\bbeta_h,\bm{\Sigma}_h]$ have prior distributions
\begin{align}\label{eq:priors}
\begin{array}{rl}
      \bbeta_h \vert \bm{\Sigma}_h &\sim \text{Matrix Normal}_{d, k}(\bB_0,\bm{\Lambda}_0^{-1},\bm{\Sigma}_h)\\
    \bm{\Sigma}_h &\sim \text{Inverse Wishart}(\bV_0,\nu_0)
\end{array}
\end{align}
with fixed hyperparameters $\bm{\eta} = [\bB_0,\bm{\Lambda}_0,\bV_0,\nu_0]$. The dimensions of the hyperparameters are: $\bB_0$ is $k\times d$, $\bm{\Lambda}_0$ is $k\times k$, $\bV_0$ is $d\times d$, and $\nu_0$ is a scalar.

\subsection{Online algorithm}\label{s:online}
The following section reviews BOCPD~\citep{Adams2007, Fearnhead2007} by making several simplifications from the state model in Section \ref{s:bayesian_linear_regression_model}. Estimating all the parameters $\btheta_h$ is burdensome and unnecessary for an online algorithm, so we marginalize them out. Given the breakpoints but marginally over $\btheta_h$, observations in different states are independent but observations in the same state are dependent via the shared parameters $\btheta_h$. The joint distribution for the data in state $h$ is

$$f(\bY_{c_{h-1}:c_{h}}|\bm{c}) = \int_{\Theta} f(\bY_{c_{h-1}:c_h}\vert \btheta_h,\bm{c})f(\btheta_h\vert \bm{\eta})d \btheta_h$$
and the joint distribution of the entire data collected through time $t$ $\bY_{1:t}$ given c is

$$f(\bY_{1:t} \vert \bm{c}) = f(\bY_{1:c_1})f(\bY_{c_1:c_2})\hdots f(\bY_{c_{G-1}:t}).$$
The form of the marginal distribution is given in Appendix A.1.

\begin{figure}
    \centering
    \subfloat{
    \includegraphics[width=0.5\linewidth, trim={0cm 2cm 0.68cm 0cm},clip]{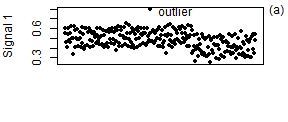}
    }
            \subfloat{
    \includegraphics[width=0.5\linewidth, trim={0cm 2cm 0.68cm 0cm},clip]{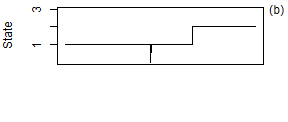}
    }\\
        \subfloat{
    \includegraphics[width=0.5\linewidth, trim={0cm 0cm 0.68cm 0cm},clip]{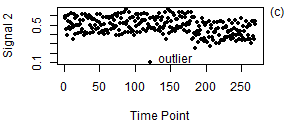}
    }
        \subfloat{
    \includegraphics[width=0.5\linewidth, trim={0cm 0cm 0.68cm 0cm},clip]{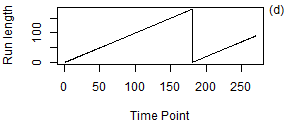}
    }
    \caption{Simulated data with changepoints $\bm{c}=\{0,181,270\}$ separating states 1 and 2. The underlying run length $r_t$ increases within each state and drops down to 0 when a new state begins at $t=181$.}
    \label{fig:rl_fig}
\end{figure}

In the monitoring paradigm, only recent state change is of interest. So we retain only the most recent changepoint, disregarding information about any previous changepoints and, consequently, the value of the current state. To keep track of the most recent changepoint, define run length $r_t$ as the number of time points that have passed since the preceding changepoint, i.e., $r_t=r$ if $g_t=g_t-r+1=g_{t-r}+1$ (see Figure \ref{fig:rl_fig}). The data associated with a particular run length $r_t=r$ is denoted $\bY_{(t-r):t}$. Let the data for each possible run length be described by the parametric model $f(\bY_{(t-r):t}\vert \btheta_t^{(r)})$ where $\btheta_t^{(r)}$ contains the model parameters associated with run length $r$ given data through time $t$. $\btheta_t^{(r)}$ is equivalent to the state specific parameter $\btheta_h \vert r=c_h-c_{h-1}$.

For online change detection, the run length posterior distribution $f(r_t\vert \bY_{1:t})$ at each time point is the quantity of interest because it contains the information about the most recent changepoint. \cite{Adams2007} formulate the following recursive message passing algorithm to calculate the run length distribution efficiently. According to Bayes' rule, the posterior run length distribution is given by:
\begin{equation}\label{eq:bayes}
    f(r_t \vert \bY_{1:t}) = \frac{f(r_t,\bY_{1:t})}{f(\bY_{1:t})}.
\end{equation}
A recursive prior distribution for the run length is defined so that the prior probability of a changepoint occurring at any time point is $\lambda$:
\begin{equation}\label{eq:runlength}
\begin{array}{l}
     f(r_t=r_{t-1}+1 \vert r_{t-1}) = 1-\lambda\\
     f(r_t=0 \vert r_{t-1}) = \lambda.
\end{array}
\end{equation}
Run length can either increment by 1, remaining in the current state, or return to 0, starting a new state, which corresponds to the interpretation that when moving forward one step in time, the number of points since the last changepoint can only increase by 1 or return to 0.

\subsection{Parameter updates}
Given the conjugate priors in (\ref{eq:priors}), the posterior distribution $f(\btheta_t^{(r)} \vert \bY_{(t-r_t):t},\bX_{(t-r_t):t})$ can be calculated exactly and the resulting posterior predictive model depends on sufficient statistics that can be updated recursively on-line. Based on (\ref{eq:likelihood}) and (\ref{eq:priors}), the posterior distribution of the run length specific parameters is
\begin{align*}
    \bbeta_t^{(r)} \vert \bY, \bX, r_t &\sim MN (\bB_t^{(r)},{\bm{\Lambda}_t^{(r)}}^{-1},\bm{\Sigma}_t^{(r)})\\
    \bm{\Sigma}_t^{(r)} &\sim IW(\bV_t^{(r)},\nu_t^{(r)})
\end{align*}
where the updated hyperparameters based on the data for $r_t=r$ are
\begin{align*}
    \bB_t^{(r)} &= (\bX_{(t-r):t}^T \bX_{(t-r):t} + \bm{\Lambda}_0)^{-1} (\bX_{(t-r):t}^T \bY_{(t-r):t} + \bm{\Lambda}_0 \bB_0)\\
    \bV_{t}^{(r)} &= \bV_0 + (\bY_{(t-r):t}-\bX_{(t-r):t} \bB_t^{(r)})^T(\bY_{(t-r):t}-\bX_{(t-r):t} \bB_t^{(r)}) + (\bB_t^{(r)}-\bB_0)^T \bm{\Lambda}_0(\bB_t^{(r)}-\bB_0)\\
    \nu_{t}^{(r)} &= \nu_0 + r\\
    \bm{\Lambda}_t^{(r)} &= \bX_{(t-r):t}^T \bX_{(t-r):t} + \bm{\Lambda}_0.
\end{align*}
Note that the updates are functions of run length specific sufficient statistics $G_t^{(r)}=\bY_{(t-r):t}^T\bY_{(t-r):t}$, $H_t^{(r)}=\bX_{(t-r):t}^T\bX_{(t-r):t}$, $K_t^{(r)} =\bX_{(t-r):t}^T\bY_{(t-r):t}$, and $r$. These statistics can be updated incrementally; e.g., $G_t^r = G_{t-1}^r+Y_tY_t^T$. Therefore, the complexity of this step does not increase with $t$.

\section{\textit{In situ} outlier detection and removal}\label{s:outlier_detection_and_removal}
In a monitoring process, outliers can be present for various reasons unrelated to the process of interest. For example, in remote sensing monitoring data, poor quality points due to clouds or shadows may occasionally slip through masking algorithms, appearing as a large spike that ends up flagged as a change. It is undesirable to flag these anomalous points as real, sustained change. The following methodology is used to identify outliers on-line and remove their effect from the analysis.

\subsection{Model uncertainty with respect to outliers}\label{s:outliers}
Suppose there are occasional outliers which, instead of following the distribution of the monitoring process in (\ref{eq:likelihood}), are in a distinct state $g_t=0$ such that
\begin{equation}\label{eq:outlier_model}
    \bY_t \vert \bm{\mu}_0, \bm{\Omega}_0, g_t=0 \sim \text{Normal}(\bm{\mu}_0,\bm{\Omega}_0),
\end{equation}
where parameters $\bm{\mu}_0$ and $\bm{\Omega}_0$ are fixed. Unlike the other states in this problem, state 0 is predefined, i.e., not learned from the data, and may be returned to at any time. The prior probability of this outlier state is $P(g_t=0)=p_0(t)$. The probability $p_0(t)$ could be based on quality control flags, but we will simply assume $p_0(t)=p_0$ for all t. Recall the motivation for keeping track of run lengths instead of all possible segmentations of the data: only recent change (or here, outliers) are of interest in the monitoring paradigm, so it is possible to greatly reduce computational complexity by keeping track of only the most recent outlier. Define $o_t \in \mathcal{O}_t =  \{\emptyset,1,\hdots,t \}$ as the number of time points since the most recent outlier in data up to time $t$ and $o_t=\emptyset$ corresponds to no outliers. The prior probability of each model being the correct model is $f(o_t=s)$.

The recognition of different possible models based on outlier inclusion parallels ideas in ~\cite{pmlr-v80-knoblauch18a} regarding BOCPD with model selection. In this case, however, the number of models under consideration increases with time, one model for each possible most recent change (and outlier). For this reason, as well as to set up later approximations, the prior probability of each outlier model $o_t$ is independent of its probability at the previous time step, $o_{t-1}$.

\subsection{Online outlier detection and removal}\label{s:online_outlier_detection_and_removal}
At time point $t$, consider models for all possible run lengths and most recent outliers. The likelihood can be simplified based on the assumption of independence between states to
\begin{equation}
    f(\bY_{1:t}\vert r_t,o_t) = f(\bY_{(t-r):(s-1)},\bY_{(s+1):t}\vert o_t=s,r_t=r)f(\bY_s\vert o_t=s)
\end{equation}
The joint distribution is, recursively,
\begin{equation}\label{eq:joint_outlier}
\begin{array}{rl}
    f(\bY_{1:t},r_t,o_t) &= \sum_{r_{t-1}}f(r_t \vert r_{t-1})f(\bY_t \vert \bY_{(t-r_t):(t-1)},o_t)f(\bY_{1:(t-1)},r_{t-1}\vert o_t)f(o_t).
    \end{array}
\end{equation}
The first term is the run length prior distribution. The second term is the predictive probability of the newest datapoint $\bY_t$ given the rest of the data associated with the same model, i.e., the data within run length $r_t$ and excluding any outliers $o_t$. This can be calculated from sufficient statistics which exclude the outlier point. For an outlier point $\bY_s$ and corresponding $\bX_{s}$ with data up to time $t$, update the run length specific sufficient statistics via:
\begin{equation}\label{eq:suff}
\begin{array}{lr}
    G_t^{(r_t,s)}=(\bY_{r_t:(s-1),(s+1):t}^T\bY_{r_t:(s-1),(s+1):t}) &= (\bY_{r_t:t}^T\bY_{r_t:t})- (\bY_s^T\bY_s)\\
       H_t^{(r_t,s)}=(\bX_{r_t:(s-1),(s+1):t}^T\bX_{r_t:(s-1),(s+1):t}) &= (\bX_{r_t:t}^T\bX_{r_t:t})- (\bX_s^T\bX_s)\\
    K_t^{(r_t,s)}=(\bY_{r_t:(s-1),(s+1):t}^T\bX_{r_t:(s-1),(s+1):t}) &= (\bY_{r_t:t}^T\bX_{r_t:t})- (\bY_{s}^T\bX_{s})
    \end{array}
\end{equation}
and recalculate the model parameters for each run length and outlier $\bm{\eta}_t^{(r_t,o_t)}$ based on these. Then the predictive probability $f(\bY_t \vert \bY_{(t-r_t):(t-1)},o_t)$ is a function of the run length/outlier specific model parameters $\bm{\eta}^{(r_t,o_t)}$. The third term is the joint distribution calculated at the previous time step, summed across all possible outlier models. The fourth term is the known prior outlier distribution.
The total evidence is then
\begin{align*}
     f(\bY_{1:t}) &= \sum_{r_t=1}^{t+1} \sum_{o_t=1}^{r_t} f(\bY_{1:t},o_t,r_t).
\end{align*}
Then the following posterior probabilities are
\begin{align}
    f(o_t,r_t \vert \bY_{1:t}) &= f(\bY_{1:t},o_t,r_t)/f(\bY_{1:t})\\
    f(o_t \vert \bY_{1:t}) &= \sum_{r_t=1}^{t+1} f(r_t,o_t \vert \bY_{1:t})  \label{eq:outlier_posterior}\\
    f(r_t \vert \bY_{1:t}) &= \sum_{o_t=1}^{r_t} f(r_t,o_t \vert \bY_{1:t}).
\end{align}

\subsection{Approximation}
Two challenges exist with implementing the on-line outlier detection procedure described above: First, it relies on the assumption of either zero or one outlier for the entire monitoring process. This is an unrealistic assumption, as multiple outlying points may occur. Second, in an online monitoring context, it is computationally expensive to calculate and store all possible outlier models and their probabilities in terms of both memory and computation. The following approximation leverages the formulas in Section \ref{s:outliers} to confirm and remove suspected outliers.

First, relax the assumption of zero or one outlier for the entire process to an assumption of zero or one outlier in a window of $L_o$ time points preceding the current time point. Next, consider only calculating outlier posterior probabilities when there is reason to suspect that an outlier has occurred. One of the biggest problems that outliers present in online monitoring is that they can be flagged as changepoints. So, we will proceed with the main BOCPD algorithm, without model uncertainty due to outliers, and only trigger the outlier model calculations when a suspected change has been flagged.

Now that the window is fixed, define the prior probability of an outlier as
\begin{equation}\label{eq:outlier_prior}
\begin{array}{rl}
      f(o_t=\emptyset) &= p_0   \\
     f(o_t=s) &= \frac{1-p_0}{L_o-1}.
\end{array}
\end{equation}
Recall that in (\ref{eq:joint_outlier}), the posterior predictive distribution $f(\bY_{t}\vert \bY_{(t-r_t):(t-1)},o_t)$ can be calculated by removing outliers from the sufficient statistics, and recalculating the distribution based on updated model parameters. The third term, the joint distribution from the prior time point conditioned on the outlier model $f(\bY_{1:(t-1)},r_{t-1}\vert o_t)$ is harder to recover retroactively since, if the outlier lies outside the current run length, its distribution is unknown. Instead, the joint distributions for the models with the most recent possible outliers within the $L_o$ window are stored for every update. Then, the full joint distribution with outlier uncertainty is available from (\ref{eq:joint_outlier}) and the posterior outlier model probabilities from (\ref{eq:outlier_posterior}).

If the posterior outlier model probability exceeds a threshold $\alpha$, the suspected outlier $o_t=s$ is removed from the analysis; $\bm{\eta}^{(r_t,s)}$ and $f(\bY_{1:t},r_{t-1}\vert o_t=s)$ are carried forward as the `true' model parameters and the joint distribution in the standard BOCPD algorithm without outlier uncertainty. In the approximation, the entire outlier checking procedure will not be triggered again until a change is suspected. While averaging over uncertainty in the outlier is done in exact analysis in Section \ref{s:online_outlier_detection_and_removal}, in the approximation, the outlier and its effect are completely removed from both the sufficient statistics and joint distribution.

\section{Extracting changepoints from run length distribution}\label{s:changepoints}
The Bayesian algorithm provides posterior probabilities of a change-point for each time point. However, a rule is necessary to convert the run length distribution to a list of the most likely change-points. One of the most simple options is to set a threshold for individual changepoints, declaring a changepoint if the posterior probability of its associated run length exceeds the threshold. A shortcoming is that the method only looks at individual points. When a change occurs, it is possible that multiple changepoint candidates may have relatively large posterior probabilities. Uncertainty about the exact time of a changepoint effectively splits the probability of each one being the changepoint into fractions. This could force the changepoint probabilities for any of the individual candidates under the threshold for detection.

Consider instead the posterior probability of the changepoint occurring between $l_0$ and $l_0+L$:
\begin{equation}
    P(r_t \in \{ l_0,\hdots,l_0+L \}\vert \bY_{1:t}) = \sum_{r=l_0}^{l_0+L} P(r_t = r \vert \bY_{1:t}).
\end{equation}
However, considering that this is an online algorithm, $l_0$ can be limited by some maximum $l_{max}$ so that we only search over $l_0\in\{0,...,l_{max}\}$. Once the window is identified, the point with maximum probability within the window is returned as a changepoint.

\section{Computational details}\label{s:comp}

\begin{figure}
    \centering
    \includegraphics[width=\textwidth]{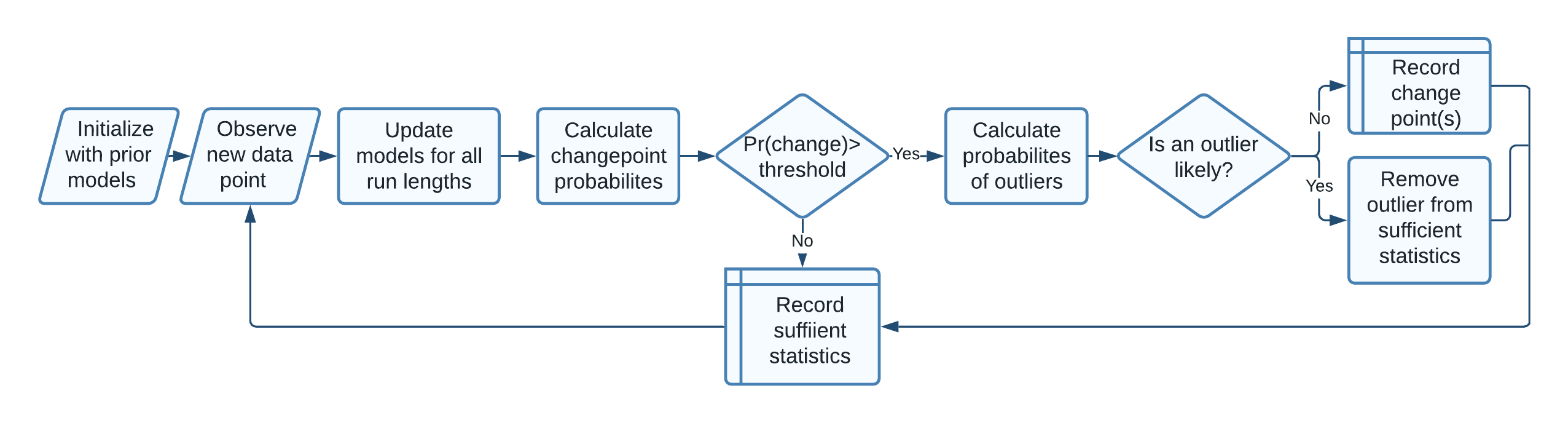}
    \caption{Workflow for BOCPD with outlier detection}
    \label{fig:workflow}
\end{figure}

The algorithms for multivariate regression BOCPD without and with outlier detection are detailed in Algorithms 1 and 2
and illustrated conceptually in Figure \ref{fig:workflow}.  To maintain computational efficiency, implementation of the online algorithm is made as lightweight as possible.  Following \citep{Adams2007}, we retain data only for run lengths with posterior probability at least $1e-4$.  We also only search over windows defined  by $L_0=20$, $L=5$ and $l_{max}=6$.

\begin{algorithm}[h]
\SetAlgoLined
 1. Initialize $\bm{\eta}_0^{(0)}=\bm{\eta}_{prior}$,
 $P(r_0=0)=1$\;
 \For{$t=1:t_{max}$}{
  2. Observe new data $\bY_t$ and record associated covariates $\bX_t$\;
  \ForEach{$r_{t}=r \in \{1,\hdots,t\}$}{
  3a. Update sufficient statistics $G_t^{(r)}$, $H_t^{(r)}$, $K_t^{(r)}$\\
  3b. update parameters for run length $r_t=r$\\
  4. Evaluate predictive probability for each possible run length:\\
  $f(\bY_t \vert \bY_{1:(t-1)},r_{t}=r) = f(\bY_t \vert \bY_{(t-r):(t-1)}) = f(\bY_t \vert \bm{\eta}_{t-1}^{(r-1)})$\\
  5. Calculate growth probabilities\\
  $f(\bY_{1:t},r_t = r=r_{t-1}+1) = f(\bY_t \vert \bY_{1:(t-1)},r_{t}=r)f(\bY_{1:(t-1)},r_{t-1}=r-1)(1-\lambda)$\\
    }
    6. Calculate changepoint probabilities\:\\
    $f(\bY_{1:t},r_t=0) = \sum_{r_{t-1}=0}^{t-1}f(\bY_t \vert \bY_{1:(t-1)},r_{t-1})f(\bY_{1:(t-1)},r_{t-1})\lambda$\\
    7. Calculate Evidence\\
    $f(\bY_{1:t}) = \sum_{r_t=0}^t f(\bY_{1:t},r_t).$\\
    8. Determine run length distribution\:\\
    $f(r_t \vert \bY_{1:t}) = \frac{f(r_t,\bY_{1:t})}{f(\bY_{1:t})}$\\
    9. Extract changepoints via threshold\\
    10. Apply Algorithm 2 for outlier detection and removal\\
    11. Update parameters for run length 0\\
    $\bm{\eta}_t^{0} = \bm{\eta}_{prior}$\\
    12. Truncate unlikely run lengths
 }
 \caption{BOCPD algorithm for multivariate linear model}
 \label{alg:bocpd}
\end{algorithm}

\begin{algorithm}[h]
\SetAlgoLined
\textbf{Input}\:\\
 $G_t^{(r_t)}$, $H_t^{(r_t)}$, $K_t^{(r_t)}$,
 $\bm{\eta}_t^{(r_t)}$,
 $f(\bY_{1:t}\vert o_t)$\\
 \For{$o_t = s \in \{(t-L_o):t\}$}{
  1. Identify outlier data $\bY_s$ and record associated covariates $\bX_s$\;
  \ForEach{$r_{t}=r \in \{1,\hdots,t\}$}{
  2a. Recover sufficient statistics\\
  $G_t^{(r,s)}$, $H_t^{(r,s)}$, $K_t^{(r,s)}$\\
  \vspace{5pt}
  2b. update parameters $\bm{\eta}^{(r,s)}$ for run length $r_t=r$, outlier $o_t=s$\\
  3. Evaluate predictive probability of the outlying point for each possible run length:\\
  $f(\bY_s \vert \bY_{1:(s-1),(s+1):t},r_{t}=r) = f(\bY_s \vert \bm{\eta}_{t}^{(r,s)})$\\
    }
 }
  4. Calculate joint distribution $f(\bY_{1:t},r_t,o_t) $ from Eq. (\ref{eq:joint_outlier})\\
 5. Calculate Evidence\\
    $f(\bY_{1:t}) = \sum_{o_t \in \mathcal{O}_t} \sum_{r_t=0}^t f(\bY_{1:t},r_t,o_t).$\\
  6. Determine outlier distribution  $f(o_t \vert \bY_{1:t})$ from Eq. (\ref{eq:outlier_posterior})\:\\
  7. Identify most likely outlier model\:\\
    $o_t^* = \underset{o_t \in \mathcal{O}_t}{max}(f(o_t \vert \bY_{1:t})>\alpha)$\\
  8. Pass on model parameters and joint distribution associated with the most likely model\:\\
       $ \bm{\eta}^{(r_t)} \leftarrow \bm{\eta}^{(r_t,o_t^*)}$\\
       $f(\bY_{1:t},r_t) \leftarrow f(\bY_{1:t},r_t \vert o_t^*)$\\
 \caption{Outlier detection and removal}
 \label{alg:outlier}
\end{algorithm}

Prior specification is a key step and should take real data and/or subject matter expertise into account when available. For example, in the data analysis of Section \ref{s:app} we fit multivariate linear regressions to a sample of historical time series from the geographical region of interest and record the estimated hyperparameters that govern the distributions of the estimated model parameters (See Appendix A.2).  However, in the simulation study of Section \ref{s:sim} we use a combination of the true underlying process priors and uninformative priors.

\section{Simulation study}\label{s:sim}

Here we apply the methods described in Sections \ref{s:multivariate_changepoint_model} and \ref{s:outlier_detection_and_removal} to simulated data to test the performance of BOCPD with multivariate regression and outlier detection versus several other monitoring algorithms.

\subsection{Data generation}\label{s:sim:gen}
A $d=2$ dimensional response $\bY_t$ is observed at each time $t\in\{1,...,270\}$. Suppose the data follows one model during a stable period from $t=0,\hdots,t_*-1$ and a new model after time $t_*=181$. The model is
\begin{align*}
    \bY_t^T &= \mu_t\mathbf{1}_2 + \bx_t^T \bm{\beta} + \be_t, \quad \be_t \iid N(0,\Sigma_t),
\end{align*}
 where $\mu_t=\mu_0=0.5$ for $t<t_*$ and $\mu_t=\mu_*$ for $t\ge t_*$ and $\Sigma_t$ is the $d\times d$ covariance matrix at time $t$. The covariates are $\bx_t^T = \begin{bmatrix}sin(2\pi t) & cos(2\pi t) & t \end{bmatrix}$ to represent two components of a seasonal trend and a linear trend.  The mean trend is either set to $\bm{\beta}=\bm{0}$ to omit seasonality or simulated as in (\ref{eq:priors}) to give seasonality. For the first eight simulation scenarios the covariance is constant over time, $\Sigma_t=\Sigma_0$, and the mean and covariance parameters in (\ref{eq:priors}) are set to
 $$
 \bm{B} = \begin{bmatrix}
    0.1 & 0.1\\
    0.04 & 0.04\\
    0 & 0
    \end{bmatrix}, \hspace{12pt}
    \bm{V} = \frac{1}{1000}\begin{bmatrix} 1 & \rho_0\\
 \rho_0 & 1\end{bmatrix},$$
    $\bm{\Lambda} = 10 \bm{I}_3$ and $\nu=20$.  For the final scenario, the covariance changes at the break point so that $\Sigma_t=\Sigma_0$ for $t<t^*$ and $\Sigma_t=\Sigma_*$ for $t\ge t^*$, where $\Sigma_0$ is drawn as before and $\Sigma_*$ is simulated as $\Sigma_0$ except with correlation parameter $\rho_*$ replacing $\rho_0$. 
In each simulation, a single outlier $\by_s=[0.8,0.1]$ is introduced at a random time $s\sim\mbox{Uniform}(90,270)$. The nine scenarios vary by the mean after the change, $\mu_*$, the error correlation before the change, $\rho_0$, the error correlation before the change, $\rho_*$, and the presence of seasonality as defined in Table~\ref{tab:sim_settings}.

\begin{table}[h]
    \caption{Simulation settings for data generation are defined by the mean after the change, $\mu_*$, correlation of the data before ($\rho_0$) and after ($\rho_*$) the changepoint, and whether seasonality is present ($\bbeta\ne \bm{0}$) or not ($\bbeta=\bm{0})$.}
    \centering
    \begin{tabular}{ccccc}
    \hline
         Case & $\mu_*$ & $\rho_0$ & $\rho_*$ & Seasonality\\
         \hline
         1 & $0.4$ & $0.0$ & - & N\\
         2 & $0.3$ & $0.0$ & - & N\\
         3 & $0.4$ & $0.9$ & - & N\\
         4 & $0.3$ & $0.9$ & - & N\\
         5 & $0.4$ & $0.0$ & - & Y\\
         6 & $0.3$ & $0.0$ & - & Y \\
         7 & $0.4$ & $0.9$ & - & Y \\
         8 & $0.3$ & $0.9$ & - & Y \\
         9 & $0.5$ & $0.5$ & $-0.5$ & Y
    \end{tabular}
    \label{tab:sim_settings}
\end{table}

\subsection{Competing methods}\label{s:sim:methods}
The proposed changepoint analysis using BOCPD with outlier detection (BOCPD-OD) is compared to continuous change detection (CCD; \citeauthor{Zhu2012a}, \citeyear{Zhu2012a}), BOCPD with multivariate regression, the multivariate formulation of BOCPD with model selection (BOCPDMS; \citeauthor{pmlr-v80-knoblauch18a}, \citeyear{pmlr-v80-knoblauch18a}) and robust BOCPD with model selection (rBOCPDMS; \citeauthor{NEURIPS2018_a3f390d8}, \citeyear{NEURIPS2018_a3f390d8}). Competing methods were limited to those that can handle multivariate responses.  Due to difficulties including seasonal trends, BOCPDMS and rBOCPDMS are run only on the first four simulation cases. Implementation details for these methods are given in Appendix A.3. The key hyperparameters/tuning parameters for the BOCPD-OD methods are: prior changepoint probability $\lambda=1/270$, prior outlier probability parameter $p_0 = 0.5$ and outlier threshold $\alpha=0.9$.  The remainder of the prior distributions for this and other methods is given in the Appendix.

\subsection{Metrics}\label{s:sim:metrics}
For evaluation, we consider change detection as a classification problem where each point is either a changepoint (CP) or not~\citep{Killick2012,Aminikhanghahi2017}. The definition of a true positive is a declared  changepoint within a tolerance of $tol=5$ of the true changepoint as in \cite{Burg2020}. In the case of multiple detections within the truth tolerance, only a single true positive is recorded. A false positive is recorded when a declared CP lies outside of the tolerance range from the truth. The latency for detection is the number of points between when the identified changepoint occurred and when it was first declared. Let ${\hat t}_j$ be the $j$th out of $J$ changepoints declared and $t_*$ be the true changepoint. Then
\begin{align*}
    TP &= I \left ( \left [\sum_{j=1}^J I\left (\lvert {\hat t}_j - t_* \rvert \le \text{tol} \right ) \right ] \geq 1 \right )\\
    FP &= \sum_{j=1}^J I\left (\lvert {\hat t}_j - t_* \rvert > \text{tol} \right ),
\end{align*}
where $I(\cdot)$ is the indicator function. The classes are unbalanced so we also consider  precision and recall.  In this simulation study, the precision is $P=TP/J$, and since there is only one true changepoint, recall is $R=TP$. The F-score of \cite{VanRijsbergen1979}, $F=2PR/(P+R)$, balances precision and recall into a single performance metric; a larger F score is preferred. In addition to the F-score, the latency of detection, or the number of points delay between the true change and when it is first declared, is recorded for each TP. Finally, the average computation times, Time (ms), for updating the model with one observation for a simulated data set are compared.

\subsection{Results}\label{s:sim:results}
The simulation results are recorded in Table \ref{tab:sim_results}. See the results in Appendix A.4 for performance metrics for different prior hyperparameters and tuning parameters that suggest the method is robust to small perturbations of these parameters. The proposed BOCPD-OD method routinely records top or near top F-scores and seldom misses a detection. CCD is the fastest method with competitive latency and performs well for cases with a larger mean shift, but has notably poorer performance for small mean shifts in terms of both $TP$ and $FP$. The lower $TP$ is expected because the three anomalous observations in a row required to signal change are less likely for a smaller shift. Increased $FP$ is likely due to incorporation of data before and after the changepoint into the model, resulting in poor model fits and, therefore, poor predictions. rBOCPDMS records competitive F-scores, but its latency as well as run time are much larger than for the other methods. In case 9, where correlation shifts and mean does not, CCD fails to detect the change as expected since it treats signals independently. BOCPD and BOCPD-OD are much better at picking up the correlation change.

\begin{singlespace}
\begin{table}[H]
\centering
\caption{Average metrics (standard errors are in subscripts) for the simulation
study. The scenarios are defined in Table 1.\\} 
\label{tab:sim_results}
\begin{tabular}{rllllll}
  \hline
Scenario & Method & TP & FP & F-score & Latency & Time (ms) \\ 
  \hline
  1 & CCD & 0.76\textsubscript{4e-04} & 0.34\textsubscript{5e-04} & 0.86\textsubscript{2e-04} & 2.00\textsubscript{0.00} & 6.26\textsubscript{7e-04} \\ 
    1 & BOCPDMS & 1.00\textsubscript{3e-05} & 1.06\textsubscript{0.001} & 0.82\textsubscript{2e-04} & 2.90\textsubscript{0.001} & 17.74\textsubscript{2e-04} \\ 
    1 & rBOCPDMS & 0.99\textsubscript{0.001} & 0.20\textsubscript{0.006} & 0.96\textsubscript{0.001} & 21.19\textsubscript{0.05} & 2709.05\textsubscript{2.31} \\ 
    1 & BOCPD & 0.94\textsubscript{2e-04} & 1.03\textsubscript{5e-04} & 0.79\textsubscript{1e-04} & 3.47\textsubscript{0.002} & 24.62\textsubscript{0.001} \\ 
    1 & BOCPD-OD & 0.99\textsubscript{1e-04} & 0.32\textsubscript{6e-04} & 0.94\textsubscript{1e-04} & 3.34\textsubscript{8e-04} & 30.00\textsubscript{0.002} \\ 
   \hline
  2 & CCD & 1.00\textsubscript{0.00} & 0.11\textsubscript{3e-04} & 0.98\textsubscript{6e-05} & 2.00\textsubscript{0.00} & 6.22\textsubscript{7e-04} \\ 
    2 & BOCPDMS & 1.00\textsubscript{0.00} & 1.29\textsubscript{0.001} & 0.79\textsubscript{2e-04} & 1.27\textsubscript{0.001} & 17.76\textsubscript{1e-04} \\ 
    2 & rBOCPDMS & 0.94\textsubscript{0.002} & 0.28\textsubscript{0.006} & 0.93\textsubscript{0.001} & 33.57\textsubscript{0.10} & 2713.88\textsubscript{1.97} \\ 
    2 & BOCPD & 1.00\textsubscript{3e-05} & 1.02\textsubscript{4e-04} & 0.80\textsubscript{7e-05} & 3.40\textsubscript{0.002} & 24.36\textsubscript{0.001} \\ 
    2 & BOCPD-OD & 1.00\textsubscript{0.00} & 0.29\textsubscript{6e-04} & 0.95\textsubscript{1e-04} & 3.29\textsubscript{0.001} & 30.14\textsubscript{0.002} \\ 
   \hline
  3 & CCD & 0.66\textsubscript{5e-04} & 0.49\textsubscript{6e-04} & 0.81\textsubscript{2e-04} & 2.00\textsubscript{0.00} & 6.27\textsubscript{7e-04} \\ 
    3 & BOCPDMS & 1.00\textsubscript{0.00} & 1.62\textsubscript{0.001} & 0.75\textsubscript{2e-04} & 2.62\textsubscript{0.002} & 17.84\textsubscript{3e-04} \\ 
    3 & rBOCPDMS & 0.98\textsubscript{0.001} & 0.37\textsubscript{0.008} & 0.93\textsubscript{0.001} & 19.68\textsubscript{0.06} & 2796.65\textsubscript{2.23} \\ 
    3 & BOCPD & 0.92\textsubscript{3e-04} & 0.91\textsubscript{3e-04} & 0.80\textsubscript{1e-04} & 3.64\textsubscript{0.002} & 24.61\textsubscript{0.002} \\ 
    3 & BOCPD-OD & 0.99\textsubscript{1e-04} & 0.04\textsubscript{2e-04} & 0.99\textsubscript{6e-05} & 3.65\textsubscript{0.002} & 30.09\textsubscript{0.002} \\ 
   \hline
  4 & CCD & 1.00\textsubscript{5e-05} & 0.16\textsubscript{4e-04} & 0.97\textsubscript{8e-05} & 2.00\textsubscript{0.00} & 6.41\textsubscript{7e-04} \\ 
    4 & BOCPDMS & 1.00\textsubscript{3e-05} & 1.76\textsubscript{0.001} & 0.73\textsubscript{2e-04} & 1.41\textsubscript{0.001} & 17.91\textsubscript{1e-04} \\ 
    4 & rBOCPDMS & 0.88\textsubscript{0.003} & 0.50\textsubscript{0.007} & 0.87\textsubscript{0.002} & 33.82\textsubscript{0.10} & 2925.90\textsubscript{2.10} \\ 
    4 & BOCPD & 1.00\textsubscript{4e-05} & 0.93\textsubscript{3e-04} & 0.81\textsubscript{5e-05} & 3.31\textsubscript{0.002} & 26.19\textsubscript{0.002} \\ 
    4 & BOCPD-OD & 1.00\textsubscript{0.00} & 0.02\textsubscript{1e-04} & 1.00\textsubscript{3e-05} & 3.09\textsubscript{6e-04} & 30.16\textsubscript{0.002} \\ 
   \hline
  5 & CCD & 0.77\textsubscript{4e-04} & 0.35\textsubscript{5e-04} & 0.86\textsubscript{2e-04} & 2.00\textsubscript{0.00} & 7.00\textsubscript{0.001} \\ 
    5 & BOCPD & 0.97\textsubscript{2e-04} & 0.92\textsubscript{3e-04} & 0.81\textsubscript{8e-05} & 3.41\textsubscript{0.002} & 26.48\textsubscript{0.002} \\ 
    5 & BOCPD-OD & 0.99\textsubscript{1e-04} & 0.18\textsubscript{4e-04} & 0.96\textsubscript{9e-05} & 3.17\textsubscript{0.001} & 32.23\textsubscript{0.002} \\ 
   \hline
  6 & CCD & 0.99\textsubscript{8e-05} & 0.12\textsubscript{3e-04} & 0.97\textsubscript{8e-05} & 2.00\textsubscript{0.00} & 7.06\textsubscript{0.001} \\ 
    6 & BOCPD & 1.00\textsubscript{0.00} & 0.92\textsubscript{3e-04} & 0.82\textsubscript{6e-05} & 3.14\textsubscript{0.001} & 26.27\textsubscript{0.002} \\ 
    6 & BOCPD-OD & 1.00\textsubscript{0.00} & 0.15\textsubscript{4e-04} & 0.97\textsubscript{7e-05} & 3.06\textsubscript{6e-04} & 32.34\textsubscript{0.002} \\ 
   \hline
  7 & CCD & 0.72\textsubscript{5e-04} & 0.45\textsubscript{6e-04} & 0.83\textsubscript{2e-04} & 2.00\textsubscript{0.00} & 6.83\textsubscript{0.001} \\ 
    7 & BOCPD & 0.87\textsubscript{3e-04} & 0.91\textsubscript{3e-04} & 0.78\textsubscript{1e-04} & 3.80\textsubscript{0.003} & 26.62\textsubscript{0.002} \\ 
    7 & BOCPD-OD & 0.95\textsubscript{2e-04} & 0.03\textsubscript{2e-04} & 0.98\textsubscript{9e-05} & 3.60\textsubscript{0.002} & 32.16\textsubscript{0.003} \\ 
   \hline
  8 & CCD & 0.98\textsubscript{1e-04} & 0.20\textsubscript{4e-04} & 0.96\textsubscript{1e-04} & 2.00\textsubscript{0.00} & 6.88\textsubscript{0.001} \\ 
    8 & BOCPD & 0.99\textsubscript{1e-04} & 0.93\textsubscript{3e-04} & 0.81\textsubscript{6e-05} & 3.38\textsubscript{0.002} & 27.83\textsubscript{0.003} \\ 
    8 & BOCPD-OD & 1.00\textsubscript{5e-05} & 0.02\textsubscript{2e-04} & 1.00\textsubscript{3e-05} & 3.07\textsubscript{5e-04} & 32.49\textsubscript{0.002} \\ 
   \hline
  9 & CCD & 0.10\textsubscript{3e-04} & 0.93\textsubscript{3e-04} & 0.55\textsubscript{1e-04} & 2.00\textsubscript{0.00} & 5.90\textsubscript{5e-04} \\ 
    9 & BOCPD & 0.68\textsubscript{5e-04} & 0.91\textsubscript{3e-04} & 0.72\textsubscript{2e-04} & 5.58\textsubscript{0.004} & 24.08\textsubscript{0.001} \\ 
    9 & BOCPD-OD & 0.81\textsubscript{4e-04} & 0.13\textsubscript{3e-04} & 0.91\textsubscript{2e-04} & 5.31\textsubscript{0.004} & 27.71\textsubscript{0.002} \\ 
   \hline
\end{tabular}
\end{table}

\end{singlespace}

\section{Land cover change using remote sensing data}\label{s:app}

In this section, we analyze the Myanmar deforestation data described in Section \ref{s:data}.  A $d=2$ dimensional response $\bY_t$ containing NDVI and SWIR2 is observed at each time $t$. Suppose $\bY_t$ follows the state distribution in (\ref{eq:likelihood}) where the covariates are
\begin{equation}
    \bX_t = \begin{bmatrix}1 & sin(\frac{2\pi t}{365}) & cos(\frac{2\pi t}{365}) & t \end{bmatrix}
\end{equation}
to account for seasonal and long term trends. For this analysis, the known stable historical time series data was taken from the geographical region of interest to set the hyperparameters of the Bayesian model $\bm{\eta}$ defined in (\ref{eq:priors}) as described in Appendix A.2. For this analysis, $L=3$ and $L_0=6$. The other hyperparameter and tuning values are the same as for the simulation study.

The multivariate regression BOCPD method both with and without outlier detection were run using the estimated hyperparameters as prior values. The monitoring methods CCD, BOCPDMS, and rBOCPDMS were run using the same values as in the simulation study. Since the BOCPDMS and rBOCPDMS software does not account for seasonality, data was preprocessed by fitting a multivariate regression and running change detection on the standardized residuals.

\begin{figure}[h]
\centering
  \subfloat[][]{\includegraphics[width=0.5\linewidth]{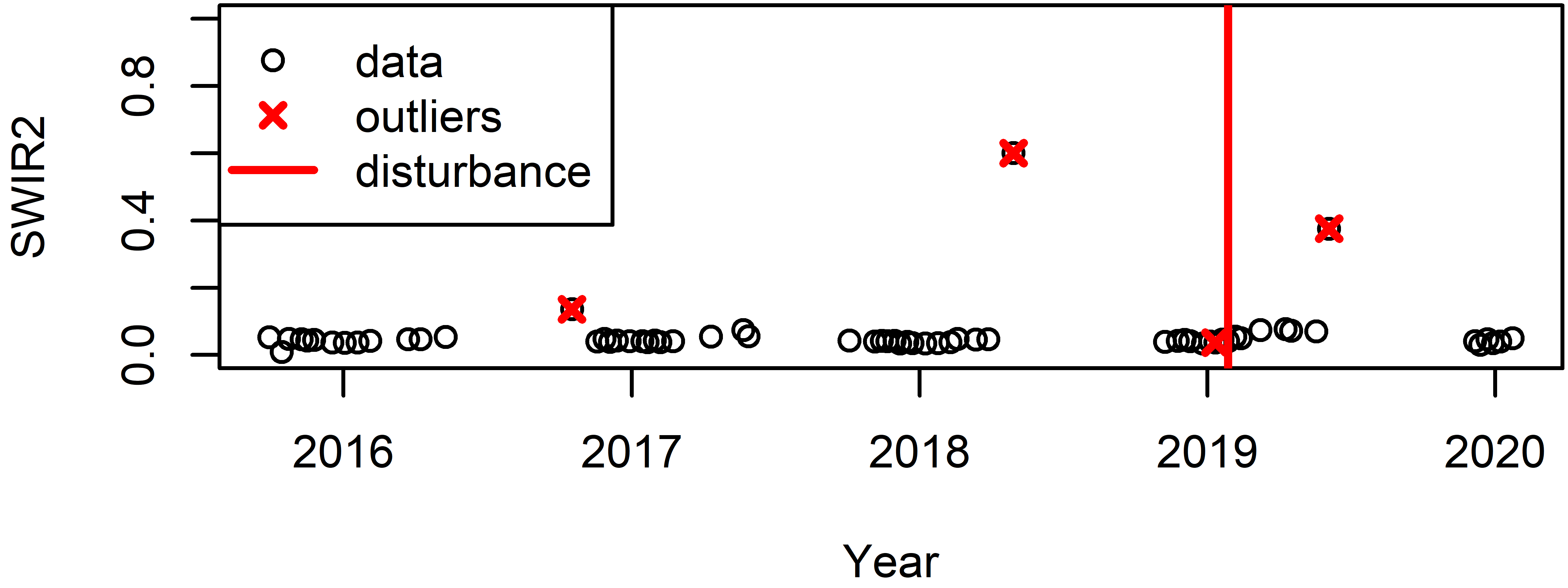}}
  \subfloat[][]{\includegraphics[width=0.5\linewidth]{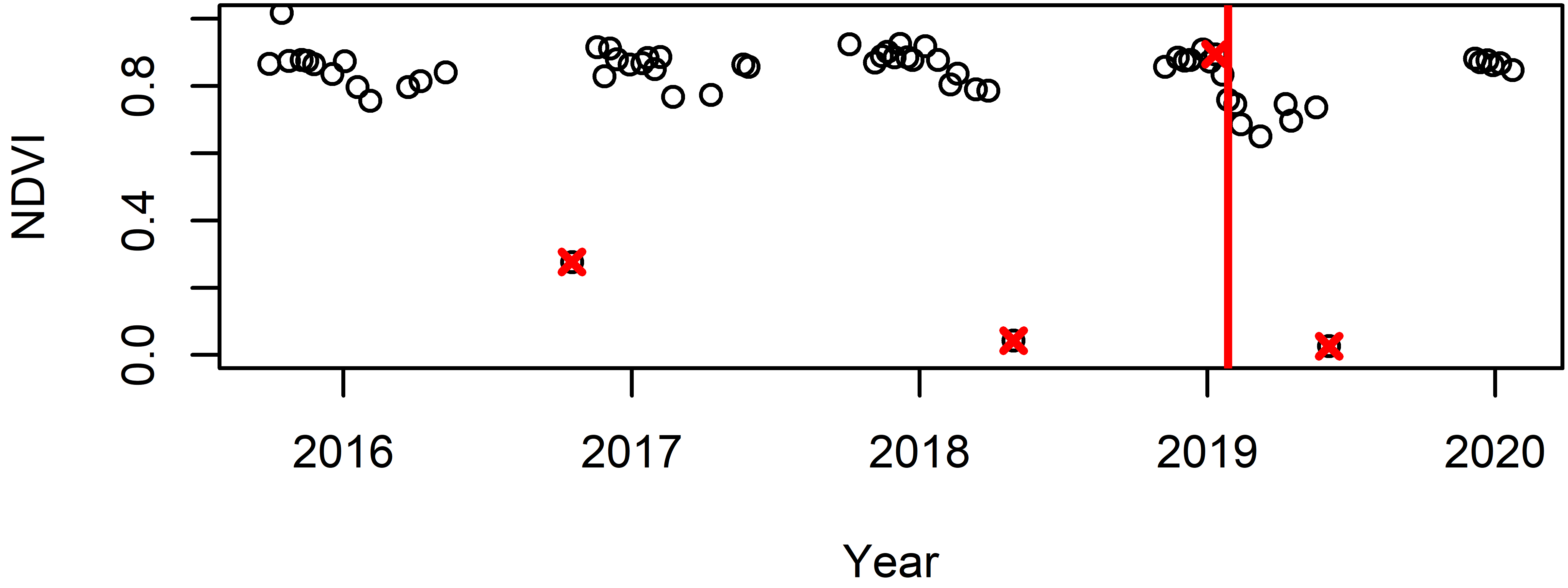}}

 \caption{Observed SWIR2 (a) and NDVI (b) with change dates and outliers detected by BOCPD-OD.}
  \label{fig:myanmar_pixel1}
\end{figure}

\begin{figure}[h]
\centering
  \subfloat[][]{\includegraphics{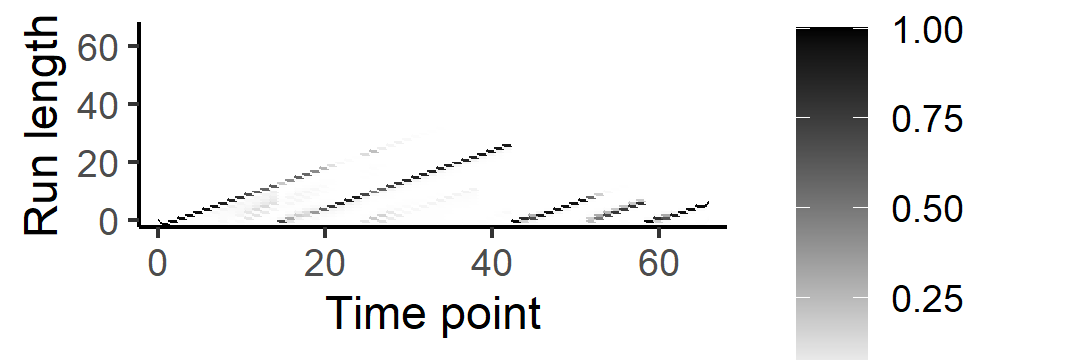}}
  \subfloat[][]{\includegraphics{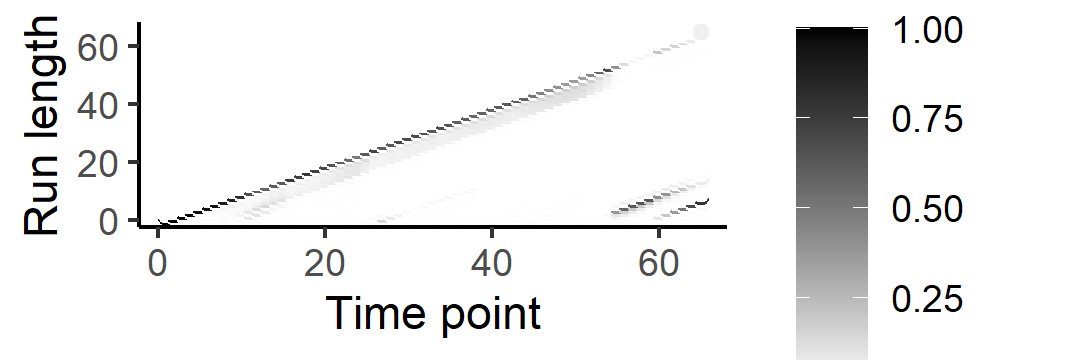}}

 \caption{The run length distribution without outlier detection (a) and with outlier detection (b). For each time point on the x axis, the probability of each run length on the y-axis being the ``correct" underlying number of points since last change is visualized by the color scale. Dark points correspond to high probability run lengths.}
  \label{fig:myanmar_rl}
\end{figure}

The detection results for a selected pixel are shown in Figure \ref{fig:myanmar_pixel1}. BOCPD-OD correctly detects the initial change date (27 January, 2019) while identifying and removing the effects of three outliers. The run length distributions for BOCPD without and with outlier detection are visualized in Figure \ref{fig:myanmar_rl}. The classic implementation of BOCPD locates numerous extraneous changepoints as shown by the probable run lengths starting over at zero repeatedly. The run length distribution for the analysis with outlier detection correctly has the most likely run length grow larger as more points are added despite anomalies. It then drops to zero around the true disturbance. Further, while the probability does not grow high enough to flag it, the run length distribution suggests that evidence is building to declare a changepoint around time point 59, where it appears that recovery from the disturbance has begun.

Since the dataset only has annotations for the first changepoint, only metrics for detections up to the annotated date of detection (plus a tolerance of 5 points as discussed in Section \ref{s:sim:metrics}) were recorded in Table \ref{tab:myanmar_results}. Further results demonstrating the sensitivity of these metrics to inputs are available in Appendix A.4. The proposed BOCPD-OD method has the best performance in terms of $TP$, $FP$, and F-score at the cost of a longer run time and a slightly larger latency than CCD. CCD excels in run time and has very good latency while remaining competitive in terms of $TP$, $FP$, and F-score. BOCPDMS and rBOCPDMS at their default values have relatively poor performance in terms of $TP$, $FP$, F-score, and time. rBOCPDMS in particular is not competitive in terms of latency, which is likely the cause of its extremely low detection rate. In many of the time series, there are not enough points recorded after the change to afford such a long latency. It is possible that tuning could improve these values, but that is beyond the scope of this comparison. Finally, we note that labels of distances described in Section \ref{s:data} and used here as the true values in fact include error and so these results should be interpreted with caution.   

\begin{singlespace}
\begin{table}[H]
\centering
\caption{Summary statistics for detection of the first change point in annotated Myanmar deforestation data.} 
\label{tab:myanmar_results}
\begin{tabular}{rlllll}
  \hline
 Metric & TP & FP & F-score & Latency & Time (ms) \\ 
   \hline
CCD & 0.85 & 0.45 & 0.88 & 2.00 & 3.98 \\ 
  BOCPDMS & 0.34 & 0.89 & 0.67 & 3.00 & 32.54 \\ 
  rBOCPDMS & 0.02 & 0.06 & 0.66 & 25.00 & 1110.64 \\ 
  BOCPD & 0.95 & 0.65 & 0.88 & 3.32 & 5.44 \\ 
  BOCPD-OD & 0.87 & 0.32 & 0.90 & 2.27 & 30.50 \\ 
  \end{tabular}
\end{table}

\end{singlespace}

\section{Discussion}\label{s:discussion}
In this paper, we define a multivariate linear regression framework for BOCPD. We introduce an \textit{in situ} outlier detection and removal procedure based on the probability of inclusion for individual datapoints to add robustness to rare outliers without sacrificing latency and while keeping run time low. BOCPD-OD elegantly models multivariate signals about a linear model trend while accounting for correlation structure. Unlike some commonly used methods in the field which require a long stable period over which to learn trends, BOCPD-OD leverages prior information from historical or geographical data to support faster learning of model parameters and, therefore, enables detection of multiple changes within a relatively short period of time.

In the remote sensing application, exploration of more regression options for the seasonality in the model, in the context of BOCPD, is open for further investigation. While NDVI and SWIR2 worked well for deforestation monitoring, other combinations of indices should be considered depending on the change type of interest. Clustered outliers can still trigger false positive detections and further research may address this rare case. Further, data quality is not necessarily a binary quantity a continuous approach to inclusion of data with varying levels of quality could better take advantage of available data.

\section*{Acknowledgements}
This research is based upon work supported in part by the Office of the Director of National Intelligence (Intelligence Advanced Research Projects Activity) via 2021-20111000006. The views and conclusions contained herein are those of the authors and should not be interpreted as necessarily representing the official policies, either expressed or implied, of ODNI, IARPA, or the U S Government. The U S Government is authorized to reproduce and distribute reprints for governmental purposes notwithstanding any copyright annotation therein. We would like to thank Ian McGregor (North Carolina State University Center for Geospatial Analytics) and Natalie Chazal (North Carolina State University Biological and Agricultural Engineering Department) for collecting and annotating the Myanmar deforestation dataset. Data courtesy of the U.S. Geological Survey.



\section*{Appendix A.1: Derivation details}\label{app:math}
\paragraph{Joint run length and data distribution}
The posterior can be rewritten so that it is a recursive function of information from the previous time step and the new data ~\citep{Adams2007}:
\begin{align*}
    f(r_t,\bY_{1:t}) &= \sum_{r_{t-1}}f(r_t,r_{t-1},\bY_{1:t}) & \text{Step 1}\\
    &= \sum_{r_{t-1}}f(r_t,\bY_{t}\vert r_{t-1}, \bY_{1:(t-1)})f(r_{t-1}, \bY_{1:(t-1)})& \text{Step 2}\\
    &= \sum_{r_{t-1}}f(r_t\vert r_{t-1},\bY_{1:(t-1)})f(\bY_{t}\vert r_{t-1},r_t,\bY_{1:{t-1}})f(r_{t-1}, \bY_{1:(t-1)}) & \text{Step 3}\\
    &=  \sum_{r_{t-1}}f(r_t\vert r_{t-1})f(\bY_{t}\vert \bY_{(t-
    r_t):(t-1)} )f(r_{t-1}, \bY_{1:(t-1)}). & \text{Step 4}
\end{align*}
In Step 1, we marginalize over the discrete run lengths $r_{t-1}\in\{0,\hdots,t-1\}$. Steps 2 and 3 each follow from the definition of conditional probability. Step 4 uses the assumption of independence of data in different states to simplify terms. The first term reduces because the incrementing of the current run length given the previous one ($r_t \vert r_{t-1}$) does not depend on $\bY_{1:(t-1)}$. In the second term, $\bY_t$ is independent of any data not in the same run length based on $r_t$. $r_t$ is determined by $r_{t-1}$ unless $r_t=0$, so we can combine the information in $r_t$ and in $\bY_{1:(t-1)}$ and instead condition on $\bY_{(t-r_t):(t-1)}$. Note that if $r_t=0$, there is no dependence on any preceding data.
The recursive run length distribution $f(r_t\vert r_{t-1})$ is defined in (\ref{eq:runlength}) and the recursive joint distribution $f(r_{t-1},\bY_{1:(t-1)})$ is available from the previous time step, so only the posterior predictive distribution remains to be calculated:
\begin{equation}
    f(\bY_t \vert \bY_{(t-r_t):(t-1)}) = \int_{\Theta}f(\bY_t\vert \btheta)\pi(\btheta = \btheta_{t-1}^{(r)}\vert \bY_{(t-r_t):(t-1)}) d\btheta.
\end{equation}
Next, the growth and changepoint probabilities are given by
\begin{align*}
    f(\bY_{1:t},r_t=r_{t-1}+1) &= f(\bY_t \vert \bY_{1:(t-1)},r_{t})f(\bY_{1:(t-1)},r_{t-1})(1-\lambda)\\
    f(\bY_{1:t},r_t=0) &= f(\bY_t \vert \bY_{1:(t-1)},r_{t})\sum_{r_{t-1}}f(\bY_{1:(t-1)},r_{t-1})\lambda
\end{align*}
and
\begin{equation}
    f(\bY_{1:t}) = \sum_{r_t}f(\bY_{1:t},r_t).
\end{equation}
Now we have all the quantities and can calculate (\ref{eq:bayes}) at time point $t$.

\paragraph{Posterior predictive distribution}
The posterior predictive distribution for this model is
\begin{align*}\label{eq:post_pred}
    f(\bY_{t+1}\vert \bY_{1:t},r_t) &= \frac{f(\bY_{1:(t+1)}\vert r_t)}{f(\bY_{1:t}\vert r_t)}\\
    &= \pi^{-d/2} \frac{\Gamma(\nu_{t+1}/2)}{\Gamma(\nu_{t}/2)}\frac{\lvert \Lambda_{t+1} \rvert }{\lvert \Lambda_t \rvert}^{-d/2}\frac{\lvert{V_{t}}^{-\nu_{t+1}/2}\rvert}{{\lvert V_{t+1}}^{-\nu_t/2} \rvert}.
\end{align*}
\paragraph{Joint distribution with outliers}
The joint distribution over both run length and possible outliers is:
\begin{equation}\label{eq:joint_outlier_full}
\begin{array}{rl}
    f(\bY_{1:t},r_t,o_t) &= \sum_{r_{t-1}}f(\bY_{1:t},r_t,r_{t-1},o_t)\\
    &= \sum_{r_{t-1}}f(\bY_t,r_t \vert \bY_{1:(t-1)},r_{t-1},o_t)f(\bY_{1:(t-1)},r_{t-1}, o_t)\\
    &= \sum_{r_{t-1}}f(r_t \vert \bY_{1:(t-1)},r_{t-1},o_t)f(\bY_t \vert r_{t-1},r_t, \bY_{1:(t-1)},o_t)f(\bY_{1:(t-1)},r_{t-1}\vert o_t)f(o_t)\\
    &= \sum_{r_{t-1}}f(r_t \vert r_{t-1})f(\bY_t \vert \bY_{(t-r_t):t-1},o_t)f(\bY_{1:(t-1)},r_{t-1}\vert o_t)f(o_t).
    \end{array}
\end{equation}

\section*{Appendix A.3: Estimating prior parameters}\label{app:priors}
Suppose several processes are available in $\bY_*$ with covariates $\bX_*$ and a multivariate regression produces the estimates $\hat{\bm{B}}_i$ and $\hat{\bm{\Sigma}}_i$ for each of the processes. Then $\hat{\nu}_0$ can be solved numerically and the rest of the hyperparameters are available as functions of the estimated parameters:
\begin{align*}
    \hat{\nu}_0 &\text{ s.t. } \frac{\partial}{\partial \nu} \mathcal{L} = 0\\
    \hat{\bm{V}}_0 &= \frac{\nu-d-1}{n}\sum_{i=1}^n \hat{\bm{\Sigma}}_i\\
    \hat{\bm{B}}_0 &= \sum_{i=1}^{n} \hat{\bm{B}_i}\\
    \widehat{\bm{\Lambda}^{-1}}_0 &= \frac{1}{nd}\sum_{i=1}^n (\bm{B}_i-\hat{\bm{B}}_0) \hat{\bm{\Sigma}}_i^{-1}(\bm{B}_i-\hat{\bm{B}}_0)^T.
\end{align*}
where
$$\frac{\partial}{\partial \nu} \mathcal{L} = \frac{n}{2}\log \lvert \bm{V}_0 \rvert - \frac{np}{2}\log 2 - \frac{1}{2}\sum_{i=1}^n \log{\hat{\bm{\Sigma}}}-\frac{n}{2}\sum_{j=1}^d \psi \left (\frac{1}{2}(\nu+1-j)) \right)$$
and $\bm{V}_0$ is replaced by its estimate $\hat{\bm{V}}_0$ and $\psi(\cdot)$ is the digamma function.

\section*{Appendix A.2: Competing methods}\label{app:settings}

\paragraph{CCD:} CCD was implemented as described in \cite{Zhu2012a} using a two term harmonic model with a mean and trend.
\paragraph{BOCPDMS:} BOCPDMS was implemented using the \texttt{Python} package \texttt{bocpdms} available on GitHub (https://github.com/alan-turing-institute/bocpdms). The analysis was run on centered and scaled data with default parameters as in~\cite{Burg2020}: $a=1$, $b=1$, $\gamma = 1$, $\lambda = 270$ and the AR(1) model.
Without preprocessing, BOCPDMS does not handle seasonality well since introduces nonstationarity withing a state. Regression of a harmonic model to remove the seasonality is possible by including exogenous variables, which is included in the BOCPDMS framework~\citep{pmlr-v80-knoblauch18a}, but is not implemented in \texttt{bocpdms}. Therefore, BOCPDMS is run only on the first four simulation cases.

\paragraph{rBOCPDMS:} rBOCPDMS was implemented using the \texttt{Python} package \texttt{rbocpdms} available on GitHub (https://github.com/alan-turing-institute/rbocpdms). The analysis was run with the same defaults as BOCPDMS as well as the default robustness parameters as in~\cite{Burg2020}:
$\alpha_{p}=0.5$, $\alpha_{rlm}=0.5$. As for BOCPDMS, rBOCPDMS is run only on the first four simulation cases.

\paragraph{BOCPD with multivariate regression:} Prior choice was guided by the true parameters of the data generating process. We set $\lambda=270$, $\nu_0=20$ and
\begin{itemize}
    \item For non-seasonal cases, $\bm{B}_0 = \begin{bmatrix}
    0.5 & 0.5\\
    0 & 0\\
    0 & 0\\
    0 & 0
    \end{bmatrix}$, and for cases with seasonality, $\bm{B}_0 = \begin{bmatrix}
    0.5 & 0.5\\
    0.1 & 0.1\\
    0.04 & 0.04\\
    0 & 0
    \end{bmatrix}$
    \item $\bm{V}_0 = (\nu_0-d-1)*0.001*\begin{bmatrix}
    1 & 0.9\\
    0.9 & 1
    \end{bmatrix}$
    \item $\bm{\Lambda}_0 = 0.01*\begin{bmatrix}
    0.1 & 0 & 0 & 0\\
    0 & 10 & 0 & 0\\
    0 & 0 & 10 & 0\\
    0 & 0& 0 & 10
    \end{bmatrix}$
\end{itemize}

\paragraph{BOCPD with multivariate regression and outlier detection:} The prior parameter choices are the same as for BOCPD without outlier detection. The outlier distribution~(\ref{eq:outlier_model}) Gaussian with mean $(0.5,0.5)^T$ and covariance $2\bm{I}_2$.

\section*{Appendix A.4: Sensitivity to priors and tuning parameters}\label{app:sensitivity}
BOCPD-OD was run on simulation 7 for various values of prior parameters and tuning parameters and the average metrics are recorded in Table \ref{tab:sim_sensitivity_results}. The $\Lambda_0$ scale controls the magnitude of the inverse variance of the regression parameters; a large scale is associated with an expectation that there is little variance among the expected model parameters (and vice versa for a small scale). The most influential input is the scale of $\Lambda_0$, which affects the $TP$, $FP$, F-score, and latency.

Similarly, metrics for BOCPD-OD with various input values applied to the Myanmar deforestation dataset (with anomalous observations represented in the data) are recorded in Table \ref{tab:myanmar_sensitivity_results}. Here, where the generating $\nu_0$ is unknown, the estimate delivers the best performance with a stronger prior, $\nu_0=10$, following closely and a weak prior strength, $\nu_0=3.01$, performs poorly. A low $\alpha$ value results in lower $TP$ than the default $\alpha=0.9$. The prior outlier probability has a relatively small effect on the metrics. The overestimate of $V_0$ scale has decreased $TP$ and $FP$ while the underestimate has increased $TP$ and $FP$; both F-scores are worse than the defaults. A higher value of $L$ delivers a slightly better F-score, but comes at the cost of increased time. The low value of $L=1$ has lower $TP$, but also a lower $FP$ and a markedly lower time than the other methods.
\begin{singlespace}
\begin{table}[H]
\centering
\caption{Average metrics (standard errors are in subscripts) for the simulation sensitivity
study applied to scenario 7.\\} 
\label{tab:sim_sensitivity_results}
\begin{tabular}{lllllllllll}
  \hline
$\mu$ & $\Lambda_0$ scale & $\nu_0$ & $\alpha$ & $p_o$ & L & TP & FP & F-score & Latency & Time (ms) \\ 
  \hline
0.5 & 1 & 20 & 0.9 & 0.5 & 5 & 0.92\textsubscript{3e-04} & 0.04\textsubscript{2e-04} & 0.97\textsubscript{1e-04} & 3.87\textsubscript{0.002} & 30.25\textsubscript{0.002} \\ 
  0.4 & 1 & 20 & 0.9 & 0.5 & 5 & 0.92\textsubscript{3e-04} & 0.03\textsubscript{2e-04} & 0.97\textsubscript{1e-04} & 3.93\textsubscript{0.003} & 30.29\textsubscript{0.002} \\ 
  0.5 & 0.01 & 20 & 0.9 & 0.5 & 5 & 0.86\textsubscript{3e-04} & 0.02\textsubscript{2e-04} & 0.95\textsubscript{1e-04} & 4.41\textsubscript{0.003} & 30.02\textsubscript{0.002} \\ 
  0.5 & 100 & 20 & 0.9 & 0.5 & 5 & 0.83\textsubscript{4e-04} & 0.07\textsubscript{3e-04} & 0.93\textsubscript{1e-04} & 5.21\textsubscript{0.004} & 29.70\textsubscript{0.002} \\ 
  0.5 & 1 & 4 & 0.9 & 0.5 & 5 & 0.95\textsubscript{2e-04} & 0.06\textsubscript{3e-04} & 0.97\textsubscript{9e-05} & 3.98\textsubscript{0.003} & 30.73\textsubscript{0.002} \\ 
  0.5 & 1 & 50 & 0.9 & 0.5 & 5 & 0.91\textsubscript{3e-04} & 0.04\textsubscript{2e-04} & 0.96\textsubscript{1e-04} & 3.91\textsubscript{0.002} & 30.58\textsubscript{0.002} \\ 
  0.5 & 1 & 20 & 0.5 & 0.5 & 5 & 0.93\textsubscript{3e-04} & 0.02\textsubscript{1e-04} & 0.97\textsubscript{1e-04} & 4.30\textsubscript{0.003} & 30.59\textsubscript{0.002} \\ 
  0.5 & 1 & 20 & 0.9 & 0.1 & 5 & 0.91\textsubscript{3e-04} & 0.02\textsubscript{2e-04} & 0.97\textsubscript{1e-04} & 4.14\textsubscript{0.003} & 30.67\textsubscript{0.002} \\ 
  0.5 & 1 & 20 & 0.9 & 0.9 & 5 & 0.92\textsubscript{3e-04} & 0.07\textsubscript{3e-04} & 0.96\textsubscript{1e-04} & 3.79\textsubscript{0.002} & 30.92\textsubscript{0.002} \\ 
  0.5 & 1 & 20 & 0.9 & 0.5 & 1 & 0.74\textsubscript{4e-04} & 0.01\textsubscript{1e-04} & 0.91\textsubscript{1e-04} & 3.91\textsubscript{0.003} & 30.64\textsubscript{0.002} \\ 
  0.5 & 1 & 20 & 0.9 & 0.5 & 3 & 0.89\textsubscript{3e-04} & 0.03\textsubscript{2e-04} & 0.96\textsubscript{1e-04} & 3.94\textsubscript{0.003} & 30.93\textsubscript{0.002} \\ 
  \end{tabular}
\end{table}

\begin{table}[H]
\centering
\caption{Summary statistics comparing sensitivity of BOCPD-OD to priors and tuning parameters of detection of the first change point in annotated Myanmar deforestation data.} 
\label{tab:myanmar_sensitivity_results}
\begin{tabular}{lllllllllll}
  \hline
$\Lambda_0$ scale & $\nu_0$ & $\alpha$ & $p_o$ & $V_0$ scale & L & TP & FP & F-score & Latency & Time (ms) \\ 
  \hline
1 &  4.82 & 0.9 & 0.5 & 1 & 3 & 0.74 & 0.32 & 0.86 & 2.46 & 27.15 \\ 
  10 &  4.82 & 0.9 & 0.5 & 1 & 3 & 0.72 & 0.33 & 0.85 & 2.32 & 27.23 \\ 
  0.1 &  4.82 & 0.9 & 0.5 & 1 & 3 & 0.76 & 0.39 & 0.86 & 2.33 & 29.19 \\ 
  0.01 &  4.82 & 0.9 & 0.5 & 1 & 3 & 0.74 & 0.46 & 0.85 & 2.21 & 29.12 \\ 
  1 &  3.01 & 0.9 & 0.5 & 1 & 3 & 0.68 & 0.43 & 0.83 & 2.54 & 27.86 \\ 
  1 & 10.00 & 0.9 & 0.5 & 1 & 3 & 0.73 & 0.29 & 0.86 & 2.33 & 29.79 \\ 
  1 &  4.82 & 0.5 & 0.5 & 1 & 3 & 0.68 & 0.25 & 0.85 & 2.67 & 23.26 \\ 
  1 &  4.82 & 0.9 & 0.1 & 1 & 3 & 0.73 & 0.29 & 0.86 & 2.49 & 28.63 \\ 
  1 &  4.82 & 0.9 & 0.9 & 1 & 3 & 0.75 & 0.38 & 0.85 & 2.33 & 31.63 \\ 
  1 &  4.82 & 0.9 & 0.5 & 0.1 & 3 & 0.75 & 0.43 & 0.85 & 2.56 & 29.65 \\ 
  1 &  4.82 & 0.9 & 0.5 & 10 & 3 & 0.63 & 0.27 & 0.83 & 2.10 & 29.21 \\ 
  1 &  4.82 & 0.9 & 0.5 & 1 & 5 & 0.76 & 0.33 & 0.87 & 2.46 & 30.55 \\ 
  1 &  4.82 & 0.9 & 0.5 & 1 & 1 & 0.65 & 0.26 & 0.84 & 2.58 & 21.54 \\ 
  \end{tabular}
\end{table}

\end{singlespace}

\FloatBarrier
\end{document}